\documentstyle[aps,eqsecnum]{revtex}

\begin{document}

%\draft

\title{\bf The optimal entropy bound and the self-energy of test objects
in the vicinity of a black hole.}

\author{Avraham E. Mayo\thanks{Electronic mail:Mayo@alf.fiz.huji.ac.il}}

\address{\it The Racah Institute of Physics, Hebrew University
of Jerusalem,\\ Givat Ram, Jerusalem 91904, Israel}

\date{\today}

\maketitle

\begin{abstract}
Recently Bekenstein and Mayo conjectured an entropy bound for
charged rotating objects. On the basis of the No-Hair principle
for black holes, they speculate that this bound cannot be improved
generically based on knowledge of other ``quantum numbers'', e.g.
baryon number, which may be borne by the object. Here we take a
first step in the proof of this conjecture. The proof make use of
a gedanken experiment in which a massive object endowed with a
scalar charge is lowered adiabatically towards a Schwarzschild's
black hole and than dropped into the black hole from some proper
distance above the horizon. Central to the proof is the intriguing
fact that the self-energy of the particle receives no contribution
from the scalar charge. Thus the energy with which the object is
assimilated consists of its gravitational energy alone. This of
course agrees with the No-scalar-Hair principle for black holes:
after the object is assimilated into the black hole, any knowledge
of the scalar field properties is lost. Using the GSL, we reach
the conclusion that the original entropy bound was not improved by
the knowledge of the scalar charge. At the end we speculate on
whether or not massive vector fields may serve in the tightening
of the entropy bound.
\end{abstract}


\section{Introduction}
A number of years ago Bekenstein proposed a universal bound on the
entropy of a macroscopic object of characteristic size $R$ and energy
$E$. This bound takes the form \cite{Bound}:
\begin{equation}
S\leq 2\pi E R/\hbar
\label{firstbound}
\end{equation}
(here and henceforth we use units with $G=c=1$). Even though the
bound was originally inferred in the framework of black holes
physics, there is independent support for its validity. It is
readily satisfied for composites of nonrelativistic particles on
account of the  fact that the entropy of the assemblage is never
far removed from the number of particles involved. Likewise, the
validity of the bound was verified directly both numerically and
analytically  for free massless quantum fields encompassed in
cavities of various shapes and topologies (see review by
Bekenstein and Schiffer \cite{Review}). Moreover, the entropy
bound was recovered by Zaslavskii from the properties of the
acceleration radiation \cite{Zaslavskii1}. With respect to
self-gravitating systems, Sorkin, Wald and Jiu \cite{SWJ} afford a
convincing indication that the entropy bound (\ref{firstbound})
holds for thermal radiation on the threshold of gravitational
collapse, while Zaslavskii \cite{Zaslavskii2} proves the bound for
a system consisting of a static black hole in equilibrium with
thermal radiation in a box.

For an object with spin $s$, electric charge $e$, maximal
characteristics length scale $R$ and proper energy $E$, Bekenstein
and Mayo conjectured an improved entropy bound \cite{BekMayo2}
\begin{equation}
S\leq 2\pi {\sqrt{E^2 R^2-s^2}-e^2/2\over\hbar},
\label{optimalbound}
\end{equation}
 a synthesis of Zaslavskii's \cite{Zaslavskii3} and Hod's \cite{Hod} bounds.

By virtue of the duality of electromagnetism, this optimal bound
can be generalized to include magnetic monopole charge $g$; one
 should merely replace $e^2\rightarrow e^2+g^2$. In addition, a deeper
 question can be set forth about the prospect of giving generic bounds
 on entropy which are tighter than the conjectured bound
 (\ref{optimalbound}) on account of the object having some
conserved "quantum numbers" apart from $e$, $g$ or $s$. The case
in point would be a tighter bound for an object with definite and
known baryon number. The aim of this work is to put forward
evidence in support of the conjecture \cite{BekMayo2} that bound
(\ref{optimalbound}), with the extension to magnetic monopole,
cannot be bettered generically. By ``generically'' we mean without
knowledge of details about the object's structure and dynamics.
When these are known, it is possible to compute, by means of
statistical mechanics, bounds on the entropy which can be small
compared to bound (\ref{firstbound}) -for example see
\cite{Numer}.  But if no such information is used, we must go to
the black hole derivation of the entropy bounds. The conjecture
that bound (\ref{optimalbound}) cannot be bettered is made
specifically for this situation.

The structure of the paper is as follows. In Sec. II we briefly
review the subject of entropy bounds as it manifest itself in the
arena of black hole physics. In addition, we put forward the
motivation for investigating the issue of entropy bounds with
regard to scalar and vector fields.  In Sec. III we begin by
writing down the action functional for a massive particle coupled
to a massless scalar field. The interaction action is chosen for
simplicity and conformal invariance. The trajectories of test
particles on the black hole spacetime background are best
characterized by their constants of the motion. In Sec. IV we
compute the energy of the particle as the constant of motion
associated with the timelike Killing vector. We thus discuss the
contribution of the scalar field to the energy of the test object
by working out in closed form the scalar potential generated by a
stationary point scalar charge in the background of a spherical
static black hole. We use a simple procedure to regulate the
potential. We find that after the regulation the scalar
self-energy vanishes. Hence the scalar interaction contributes
nought to the energetics of the process. This signifies that the
object is assimilated with its gravitational energy only, {\it
e.g.} if the process is an adiabatic one, then at the end we are
left with a new spherical static black hole, whose mass is equal
to the mass of the initial black hole augmented only by the
gravitational energy of the object. The increase of the black hole
mass obviously increases its horizon's surface area. In Sec. V we
analyze the corrections to the area formula and find that the
corrections must vanish in linear theory. The change in horizon
area that results from the lowering of the object onto the horizon
furnishes a derivation from the GSL of the entropy bound
(\ref{firstbound}). We conclude by illustrating the application of
the method used here to another problem: an entropy bound for a
particle coupled to a massive vector field with a vanishingly
small but non zero mass.

\section{Entropy bounds in the framework of black hole physics:
the state of the art}

In its original form, the entropy bound (\ref{firstbound}) is
saturated by the Schwarzschild black hole. This prompted the
observation\cite{Bound} that the Schwarzschild black hole is the
most entropic object for given size and energy.  But the Kerr
black hole's entropy falls below bound (\ref{firstbound}) (this
will be true for any reasonable interpretation of the radius $R$
of the nonspherical Kerr hole).

This asymmetric state of affairs motivated Hod\cite{Hod} to search
for a tighter bound on entropy for objects with angular momentum
which is saturated by the Kerr hole.  Hod repeats Bekenstein's
derivation\cite{BHentropy,Brazil} of the minimal increment in
Kerr-Newman (KN) horizon area that is caused by an object's
infall. That derivation applied the idea of
Christodoulou\cite{Christodoulou} together with
Carter's\cite{Carter,MTW} integrals of the Lorentz equations of
motion to a particle of rest mass $\mu$ and radius $R$ moving in a
KN background. The minimal growth in horizon area was deduced from
the conservation laws and the relation they establish between the
change in black hole parameters and the energy and orbital angular
momentum of a particle in an orbit, such that the particle's center
of mass can get to distance $R$ from the horizon.
Remarkably, it turns out that the minimal area growth is independent of
the black hole parameters.  Because $\mu$ can be identified with
the total proper energy of the object, bound (\ref{firstbound})
follows from the minimal area growth and the GSL.

The  particle's spin was not taken into account in Carter's
integrals.  Hod refers instead to Hojman and Hojman's\cite{Hojman}
integrals of motion for a neutral object with spin $s$ moving on a
KN background. Repeating the argument that led to the original entropy
bound (\ref{firstbound}) and appealing to the GSL allows Hod to infer the
entropy bound (\ref{optimalbound}) with $e=0$.

Recently many researches closed in on the derivation of
Zaslavskii's \cite{Zaslavskii3} proposed bound for charged
objects. Those derivations focused on absorption of the relevant
object by a black hole, and on the concomitant change in horizon
area. Hod \cite{Hod1} makes use of the thermodynamics of a
Schwarzschild black hole, while Bekenstein and Mayo
\cite{BekMayo2} makes use of the thermodynamics of a
Reissner--Nordstr\"om (RN) black hole. Linet \cite{Linet2}
utilizes the thermodynamics of a KN black hole, with similar
results. At the center of the derivation lies the fact that  a
charged particle in a black hole's vicinity is affected  by not
only the Lorentz force from the black hole's electromagnetic field
and by the Abraham--Lorentz--Dirac radiation reaction force, but
also by the force originating from the black hole's polarization
by the particle's electric field.  Now it is known that a particle
at rest in a static black hole background does not radiate
(despite its being accelerated). For that reason, one can expect
the radiation reaction force to vanish. This suggests that we
should focus on an object lowered slowly from a large distance to
the horizon. Under this circumstances, it is possible to suppose
that only the gravitational, Coulomb and polarization forces act
upon it. This approach allows the authors mentioned above to
derive Zaslavskii's bound by use of the GSL.  Now, {\it if\/}, as
it is sometimes claimed, the GSL functions independently of
entropy bounds, there should not have been reason for this unusual
effect (black hole polarization) to supplies precisely the missing
element in the derivation of the entropy bound for charged objects
from the GSL.  This is yet another demonstration that the GSL
provides a valid road to entropy bounds.

A useful by-product of the mentioned chain of generalizations is
the revelation that the entropy bound is independent of the type
of black hole employed in the calculations. Thus, it seems that
one may choose the Schwarzschild spacetime as the simplest
settings for the study of entropy bounds.

Bound (\ref{optimalbound}) has one more merit;
any KN black hole (mass $m$, charge $q$ and angular momentum $j$)
saturates it.  The horizon area of such a black hole is\cite{MTW}
\begin{equation}
A = 4\pi(r_+^2+j^2/m^2);\qquad r_+\equiv m+(m^2-j^2/m^2-q^2)^{1/2}
\label{area}
\end{equation}
Substituting $r_+$, squaring as required, and cancelling terms
gives
\begin{equation}
A = 2\pi(4mr_+-q^2) =  2\pi[4m((r_+^2+j^2/m^2)-j^2/m^2)^{1/2}-q^2]
\label{next}
\end{equation}
In light of Eq.(\ref{area}) it is reasonable to interpret
$(r_+^2+j^2/m^2)^{1/2}$ as the radius $R$ of the hole.
Incorporating this in the last equation and dividing by $4\hbar$
gives for the black hole entropy
\begin{equation}
S_{\rm BH} ={2\pi\over \hbar}[(m^2R^2-j^2)^{1/2}-q^2/2]
\label{entropy}
\end{equation}
If we identify $m\leftrightarrow E$, $q\leftrightarrow e$ and
$j\leftrightarrow s$, this is exactly the upper bound of
Eq.(\ref{optimalbound}).  Hence the KN black hole saturates the
proposed entropy bound. This property would be lost if the bound
were modified. Hence we adopt it in the given form.

Finally, we arrive at the principal issue we hope to elucidate in this
work.

\medskip

{\bf Question}: {\it
Is it possible to improve bound (\ref{optimalbound}) generically based on  the
knowledge of other ``quantum numbers'' which may be borne by the
  object? }

\medskip

The No-Hair conjecture is central to our reasoning.  A large
amount of work has certified that a stationary black hole can have
just a few parameters. The irrefutable ones are mass, charge,
magnetic monopole and angular momentum.  Skyrmion number is a
possible addition \cite{skyrmion}, but one whose physical
significance is unclear\cite{Brazil}. Other candidates
\cite{color,BBM,Greene} are associated with unstable black
holes\cite{Brodbeck}. The technique we propose to use is
perturbative in nature. So it stand to reason to focus only on
black holes which remain stable despite the outside perturbations.
On that account, we concentrate on the KN black holes with
parameters $m$, $q$, $g$ and $j$.

In \cite{BekMayo2}, Bekenstein and Mayo give arguments against the
prospect of obtaining generic bounds on entropy, which are tighter
than bound (\ref{optimalbound}) on account of the object having an
extra additive conserved global quantity such as baryon or lepton
numbers. Also excluded are short range fields, such as the short
range, W-boson mediated, weak force. The third and last case
considered is when the extra additive quantity carried by the
object, $b$ is the source of a long--range field schematically
denoted by ${\cal B}$. For example, ${\cal B}$ can be a scalar
field with presumably small or zero mass, or a massive vector
field with vanishingly small but nonzero mass. The range may be
finite but large in comparison to the size of a typical object.
For a massive scalar or vector field this means, that on the one
hand the Compton wavelength of the object itself must be very
small on the scale of the hole, and on the other the scale of the
hole must be small compared with the range of the field. This is
granted by the smallness of the field's mass. Now the area formula
may differ from the usual area formula for a KN black hole by
terms depending on $b$, because of the perturbation that ${\cal
B}$'s energy-momentum tensor exerts on the metric. Unless ${\cal
B}$ is a gauge field which remains massless in the classical (or
low energy) limit, we cannot rule out such dependence. This is
because Birkhoff--type theorems exist only for massless vector
fields, and from our point of view, the electromagnetic field is
the only one. Thus, while the area stays constant during the
descent as required by the adiabatic theorem, $m$ may change by a
quantity of $O(b^2)$. The sign of this quantity is vague without a
specific model. Likewise, the energy with which the object is
assimilated into the black hole, will presumably have a term of
$O(b^2)$. Indeed, as before this term may be positive here. But it
does not follow that the effect of $b$ is to subdue the growth of
the horizon area. This is because the indefinite sign of the
correction to the area formula. Without calculating linear
corrections to the metric, one cannot judge whether the change in
area is incremented or depressed by $b$'s presence.

In the following sections these questions will be precisely phrased
and answered.

\section{Action  of a massive particle coupled to a conformal scalar field.}

We use the signature $\{-,+,+,+\}$ and denote the timelike
coordinate outside the black hole, presumed to be a spherical
static one, by $x^0$. The simplest parameter independent action
functional for a particle of rest mass $\mu$ coupled to a
conformal massless scalar field $\Phi$ is
\begin{equation}
S =-\int(\mu +b\Phi)\sqrt{-g_{\alpha\beta}{dx^\alpha\over
d\lambda}{dx^\beta\over d\lambda}}d\lambda
 \label{action}
\end{equation}
where $b$, a constant, is the coupling strength, $\lambda$ is a
parameter, and $x^\alpha(\lambda)$ is the trajectory of the
particle. The term proportional to $\mu$ is the action for a free
particle; that proportional to $b$ is the interaction action. The
interaction chosen here is the most natural one in that the source
term it generates for the wave equation for $\Phi$ is independent
of $\Phi$ itself. Furthermore, the interaction action is invariant
under the conformal transformation $g_{\mu\nu}\rightarrow\Omega^2
g_{\mu\nu}$, $\Phi\rightarrow\Phi\Omega^{-1}$, where $\Omega$ is
an arbitrary function. Since the free field action possesses the
same invariance \cite{parker} it follows that the wave equation
with a source will be conformally invariant. Thus, the coupling
envisaged here, apart from being the simplest one, is singled out
by its invariance properties. The same cannot be said about the
once popular derivative coupling
($\propto\,\Phi_{,\alpha}\,dx^\alpha/d\lambda$) \cite{schweber}
which is known to be unphysical on grounds of lack of
renormalizability.

In a sense, the coupling as chosen in (\ref{action}) is the analog
of minimal coupling in electromagnetism. However, it should be
noted that the analogy with the electromagnetic case stops here;
conservation of charge is not obligatory for consistency of the
Klein--Gordon equation, so there is no obstacle to permitting $b$
to vary. However, here we suppose throughout that $b$ is constant.

Variation of $S$ with respect to $x^\nu$ gives the equation of
motion
\begin{equation}
(\mu+b\Phi){D^2 x^\nu \over d\lambda^2}= \left\{{1\over
2}(\mu+b\Phi) {d\over
d\lambda}\ln\Xi-b\Phi_{,\alpha}{dx^\alpha\over
d\lambda}\right\}{dx^\nu\over d\lambda}-b\Xi\,\Phi^{,\nu}
\label{motion}
\end{equation}
where $\Xi=-g_{\alpha\beta}{dx^\alpha\over d\lambda}{dx^\beta\over
d\lambda}$. Since the action $S$ is invariant under a change of
the parameter $\lambda$, we are at liberty to impose a condition
on $\Xi$ to fix the choice of $\lambda$. Two choices are of
interest here. If we set $\Xi=1$, $\lambda$ becomes proper time
$\tau$, and Eq.~(\ref{motion}) takes the form
\begin{equation}
(\mu+b\Phi){D^2 x^\nu \over d\tau^2}= -b\left\{\Phi^{,\nu} +
\Phi_{,\alpha}{dx^\alpha\over d\tau}{dx^\nu\over d\tau}\right\}.
\label{motion1}
\end{equation}
A second useful choice is $\Xi=(1+b\Phi/\mu)^2$ which makes the
coefficient of ${dx^\nu\over d\lambda}$ in Eq.~(\ref{motion}) to
vanish. In this case $\lambda$ is just an affine parameter. The
equation of motion is
\begin{equation}\label{motion2}
  {D^2 x^\nu \over
  d\lambda^2}=-{b\over\mu}(1+{b\over\mu}\Phi)\Phi^{,\nu}.
\end{equation}
In addition we have
\begin{equation}
-g_{\alpha\beta}{dx^\alpha\over d\lambda}{dx^\beta\over
d\lambda}=\left({d\tau\over d\lambda}\right)^2
=(1+{b\over\mu}\Phi)^2
\label{affine_proper}
\end{equation}
which shows that in regions of weak $\Phi$, $\lambda$ is
essentially proper time.

\section{The energetics of the process is unaffected by the scalar field}

Consider now a test scalar charge moving in the background of
black hole spacetime which possesses a symmetry represented by a
Killing vector $\xi^\alpha$. So $\xi_{(\alpha ; \beta)}=0$ and
$\xi^\alpha\Phi_{,\alpha}=0$. The scalar-charge trajectory,
that of a particle obeying the equation of motion
Eq.~(\ref{motion}), is best characterized by its constants of
motion. It is easy to show from Eq.~(\ref{motion1}) and Eq.~(\ref{motion2})
that
\begin{eqnarray}
&&  {\cal E}\equiv-\xi_\alpha{dx^\alpha\over d\tau}(\mu+b\Phi)
\nonumber \\ && E\equiv-\xi_\alpha{dx^\alpha\over d\lambda}
\label{energy}
\end{eqnarray}
are constants of the motion in the proper time parameterizations
and in the affine parameter parameterizations respectively . Note
that ${\cal E}=\mu E$. The stationarity of the envisaged
background fixes the form of the  timelike Killing vector to be
$\xi^\alpha=\delta_t^\alpha$ for which ${\cal E}$ reduces to
\begin{equation}
{\cal E}= -(\mu+b\Phi) g_{0\beta}\,{dx^\beta\over d\tau}.
 \label{energy1}
\end{equation}
This corresponds to the usual notion of energy as measured at
infinity. Its first term expands to $\mu +{1\over 2} \mu (d{\bf
x}/dt)^2$ in the Newtonian limit. The second term, $-b \Phi$, is
thus the scalar potential energy.

In our gedanken experiment the object of rest mass $\mu$ and
scalar charge $b$, idealized as spherically symmetric, is
suspended by some means to keep it from falling freely, and is
slowly lowered radially towards the black hole.  Of course, the
forces restraining its fall change its energy measured at infinity
as it is lowered. The idea is to bring the object as close to the
horizon as possible, and then drop it in, inferring from the
energy measured at infinity at its last prefall position the
increase in horizon area that this causes. A complication - the
Unruh--Wald buoyancy in acceleration radiation\cite{GSL,UW1} - may
cause the object to float neutrally some distance from the
horizon, thus arresting the contemplated descent. But as
demonstrated by Bekenstein \cite{Newbound,PW,AR}, provided the
number of relevant particle species in nature is not large (which
seems to be true in our universe), and provided the object is
macroscopic and composed of parts that obey quantum mechanics, the
buoyancy is negligible all the way to very near the horizon, and
makes no practical difference to the energy budget of the process.

Correct to $O(b^2)$ the metric may be taken as Schwarzschild's. In
isotropic coordinates it is
\begin{equation}
ds^2 = - \left({1-{m\over 2\rho}\over 1+{m\over
2\rho}}\right)^2(dx^0)^2 + (1+{m\over 2\rho})^4 \left[d\rho^2 +
\rho^2 (d\theta^2 +\sin^2 \theta\, d\phi^2)\right]
\label{metric}
\end{equation}
We see that the horizon resides at $\rho=m/2$. Because the object
is nearly stationary, its 4-velocity, which we normalize to $-1$,
must have the form $dx^\alpha/d\tau\approx
\{(-g_{00})^{-1/2},0,0,0\}$. Substitution in Eq.(\ref{energy1})
from the metric gives for the energy, when the object's CM is at
$\rho=a$ and $\theta=0$,
\begin{equation}
{\cal E} = \left(1-{m\over 2a}\right)\, \left(1+{m\over
2a}\right)^{-1}(\mu+b\Phi)_{\rho=a, \theta=0} \label{energy2}
\end{equation}

As elucidated by Vilenkin\cite{Vilenkin} and corroborated by Smith
and Will\cite{Smith}, by contrast to the situation in flat
spacetime, in the presence of a Schwarzschild black hole the
self--energy of an electric charge measured at infinity is
modified. As we intend to show, this is surprisingly {\it not} so
for the scalar field.

Eq.(\ref{Final_Phi}) of the Appendix gives $\Phi(\rho,\theta)$,
the scalar potential due to a stationary (or nearly so) point
scalar charge of strength $b$ in the background of a spherical
black hole. This expression, accurate to $O(b^2)$, is the scalar
analog of an early brilliant solution by Copson\cite{Copson} who
found the electric field resulting from a charge in the
Schwarzschild background. The solution (\ref{Final_Phi}) was also
found independently by Linet\cite{Linet}.

$\Phi(\rho,\theta)$ naturally diverges at $\rho=a$ and $\theta=0$,
the position of the scalar charge. Thus if we want to use it for
our finite object, we must regulate the potential before going to
the limit $\rho\rightarrow a$ and $\theta\rightarrow 0$ as
required by formula (\ref{energy2}).

The simplest procedure is as follows (compare with
\cite{BekMayo2}). We reexpress $\Phi$ in terms of new coordinates
$\{\varrho, \vartheta,\phi\}$ centered on the charge, rather than
on the black hole center, as was the case for $\{\rho,
\theta,\phi\}$, but sharing the same polar axis. This implies the
substitutions
\begin{eqnarray}
&& \rho \cos\theta \rightarrow a + \varrho \cos\vartheta,
\nonumber \\ && \rho\rightarrow
\sqrt{a^2+\varrho^2+2a\varrho\cos\vartheta}.
\end{eqnarray}
A small metric sphere of proper radius $R$ located at
$\{\rho,\theta\}=\{a,0\}$  is the coordinate sphere
$\varrho=(1+{m\over 2a})^{-2}R; \forall\vartheta$. Since $\varrho$
is the coordinate distance from the charge it makes sense to
expand $\Phi$ in a Laurent series in $\varrho$:
\begin{equation}
\Phi=-{b\over\varrho (1+{m\over 2a})^2}-{b\over a (1+{m\over 2a})^2}\,
{\cos\,\vartheta\over (2a/m)^2-1}+O(\varrho).
\label{Phi}
\end{equation}

The divergent term in Eq.(\ref{Phi}) corresponds to the scalar
potential of the charge $b$ in flat spacetime; there we expect
$\Phi=-b \varrho^{-1}$. A factor $ (1+{m\over 2a})^{-2}$ is required
to convert the coordinate distance $\varrho$ to proper distance. Thus
when taking the limit $\rho\rightarrow a$ and $\theta\rightarrow 0$
($\varrho\rightarrow 0$) of $\Phi$, we must discard the first term in
the r.h.s. of Eq.(\ref{Phi}).

Our spherically symmetric finite object samples all directions
about its center without discrimination.  Because the metric also
looks isotropic in coordinates $\{\varrho, \vartheta, \phi\}$, we
must thus average out the second term in the r.h.s. of
Eq.(\ref{Phi}) over all angles $\vartheta$ and $\phi$; as a result
its contribution vanishes. Terms of $O(\varrho)$ vanish as the
size of the object shrinks.  Thus the entire scalar contribution
to ${\cal E}$ vanishes! This result was also found independently
by Zelnikov and Frolov \cite{Zelniko}, Wiseman \cite{Wiseman} and
Burko \cite{Burko}. Substituting this in Eq.(\ref{energy2}) we
find
\begin{equation}
{\cal E}=\mu\left(1-{m\over 2a}\right)\, \left(1+{m\over
2a}\right)^{-1} + O\left({b^3\over m^2}\right)
 \label{potentialenergy}
\end{equation}
where we have included the next higher order correction to the
energy due to the coupling to the scalar field.

 When the object is near the horizon, the proper distance
from its CM to the horizon is
\begin{equation}
\ell \equiv \int_{m/2}^a (g_{\rho\rho})^{1/2} d\rho \approx
4(a-{m/2}) +O[(a-m/2)^2].
 \label{ell}
\end{equation}
Expressing $a$ in Eq.~(\ref{potentialenergy})  in terms of
$\ell$ by means of Eq.(\ref{ell}) we get
\begin{equation}
{\cal E} =\left({\mu\ell\over 4m}\right) \left[1+O\left({\ell\over
m}\right)\right] + O\left({b^3\over m^2}\right).
 \label{energy3}
\end{equation}
Corrections of $O({\ell/ m})$ are duly neglected, as are those of
$O({b^3/m^2})$ by virtue of the assumed smallness of $\ell$ and
$b$ compared to the large mass of the black hole.  The gradual
approach to the horizon must stop when the proper distance from
the object's CM to the horizon reaches the object's radius $R$.
Hence,
\begin{equation}
{\cal E}\geq {\mu R\over 4m}.
\label{energyfinal}
\end{equation}

\section{The area formula and the entropy bound}

As mentioned, our primary concern is with changes in the horizon
area. The area formula must be corrected for the perturbation of
the metric originating in the object, which in linear approximation
should be of $O(\mu)$ and $O(b^2)$, the first
caused by the energy momentum tensor of the object's mass, and the
second by the energy momentum tensor of the scalar field. We now
argue  that the corrections to the area formula actually appear only
in the next higher order.

First suppose the area $A$ was in fact perturbed in linear
approximation to $O(\mu)$ and $O(b^2)$. By spherical symmetry of
the background these corrections would not depend on the direction
along which the object was lowered. If $N$ equal bodies were
lowered, each along a different radial direction, the perturbation
would be $N$ times larger by linearity of the approximation.  But
if enough bodies were disposed on a spherical shell concentric
with the black hole, the metric perturbation due to the energy
momentum tensor of {\it  the object's mass} at the horizon should
tend to zero by Birkhoff's theorem \cite{MTW} that the metric
exterior to a spherical black hole is exactly Schwarzschild if the
surroundings are spherically symmetric. We thus get a
contradiction unless we admit that the perturbation of $O(\mu)$
vanish in linear theory.  Any corrections to $A$ must thus be of
higher order, like $O(\mu^2)$, {\it etc.\/}

What about the {\it scalar} perturbation? Since there is no analog
of Birkhoff's theorem for scalar fields, we must verify directly
that the perturbation to the horizon area formula is of order
higher than $O(b^2)$.

As before we assume that  a large number of scalar charges are
disposed on a spherical shell concentric with the black hole. By
the linearity of the approximation the perturbation to the metric
from the scalar charge's field should be of $O(b^2)$.  Now, a
spherically symmetric perturbation to a static spacetime
surrounding a black hole can be expressed by
\begin{eqnarray}
&& g_{\mu\,\nu}\approx g^{(B)}_{\mu\,\nu} +h_{\mu\,\nu}, \nonumber
\\ &&
h_{\mu\,\nu}\,dx^\mu\,dx^\nu=-u(\rho)\,{\left (dx^0\right )}^2+
f(\rho)(d\rho^2+\rho^2
d\Omega^2)
\label{perturbation}
\end{eqnarray}
where $g^{(B)}_{\mu\,\nu}$ is the background metric and
$d\Omega^2$ is the background line element on the 2-sphere,
$d\Omega^2=d\theta^2+\sin\,\theta^2 d\phi^2$.

The horizon area formula is
\begin{equation}
A=\int \,\sqrt{-g}_{\rho=\rho_{\cal H}}\,d\theta d\phi
\label{A_correct}
\end{equation}
where $\sqrt{-g}$ and $\rho_{\cal H}$ are to be evaluated at the
(perturbed) location of the horizon. The correction to $A$ in
linear theory can be determined by the following procedure. To
first approximation, the volume element is
\begin{equation}
 g = det \, (g_{\mu\nu})= \epsilon^{\alpha\beta\gamma\delta}\,
g_{0\alpha}g_{\rho\beta}g_{\theta\gamma}g_{\phi\delta}=
g^{(B)}(1+h) + O(b^4)
\label{det}
\end{equation}
where $\epsilon^{\alpha\beta\gamma\delta}$ is the Levi-Civita
tensor and $h=g^{(B)}_{\alpha\beta}\, h^{\alpha\beta}$. Terms of
$O(b^4)$ are to be understood as quadratic in $h$. Using
Eq.~(\ref{A_correct}) for the area formula, we find
\begin{equation}
A= A^{(B)}\left(1+{1\over 2}\int h_{\rho=\rho_{\cal H}} \,d\theta
d\phi\right)+O(b^4).
\label{A}
\end{equation}

Where does the new horizon resides? The point $\rho=\rho_{\cal H}$
where $g_{00}$ vanishes is to be interpreted as the location of
the horizon (if several zeros exist the location of the horizon
corresponds to the outermost one). Therefore, correct to $O(b^4)$
the new horizon resides at
\begin{equation}
\rho_{\cal H}=\rho^{(B)}_{\cal H}+\delta\rho =\rho^{(B)}_{\cal
H}-\left .{h_{00}\over (g^{(B)}_{00})'}\right |_{\rho
=\rho^{(B)}_{\cal H}}+O(b^4). \label{horizon}
\end{equation}
 Hence, correct to $O(b^2)$, $h$ in Eq.~(\ref{A}) can be
evaluated at $\rho=\rho^{(B)}_{\cal H}$.

Now, the field equation for $h_{\mu\nu}$ are as follows
\cite{MTW}. First the Ricci tensor due to the perturbation is
\begin{eqnarray}
&& R_{\alpha\beta}=R^{(B)}_{\alpha\beta}+R^{(1)}_{\alpha\beta}(h)
+O(b^4), \nonumber \\ && R^{(1)}_{\alpha\beta}(h)={1\over 2}
\left(-h_{;\alpha\beta}-{h_{\alpha\beta ;\gamma}}^\gamma
+{h_{\gamma\alpha ;\beta}}^\gamma+ {h_{\gamma\beta
;\alpha}}^\gamma\right) \label{h_Eq}
\end{eqnarray}
where $R^{(B)}_{\alpha\beta}$ is the background Ricci tensor,
which vanishes for the Schwarzschild spacetime. Next, a useful
identity can be established
\begin{equation}
R={h^{\alpha\beta}}_{;\alpha\beta}-{h_;\,^\beta}_\beta\, +O(b^4).
\label{ricci}
\end{equation}
From the trace of Einstein's equations, $R=-8\pi T$, where
$T=-\Phi_{,\alpha}\Phi^{,\alpha}$ is the trace of the energy
momentum tensor of the massless spherically symmetric scalar
field. It is an invariant of the geometry. By Eq.~(\ref{general})
of the Appendix with charge distribution density appropriate for a
spherical symmetric configuration of scalar charges, $T$ is
$O(b^2)$. The invariance of $T$ and hence of $R$ signifies that
their values at the perturbed horizon (as given in
Eq.~(\ref{horizon})) which is by all means a physically regular
surface, must both be finite. Else their divergence would give
rise to curvature singularity at the horizon, a thing that would
render our perturbation approach invalid. Taylor expanding
$R^{(1)}$ around $\rho^{(B)}$ we find
\begin{equation}
R^{(1)}(\rho^{(B)}_{\cal H}+\delta\rho)= R^{(1)}(\rho^{(B)}_{\cal H})+
(R^{(1)}(\rho^{(B)}_{\cal H}))'\,\delta\rho +O(b^6).
\end{equation}
The second term in the r.h.s. of the equation is obviously
 $O(b^4)$ (see Eq.~(\ref{horizon})), hence it can not cancel any
  divergence due to the first term in the r.h.s. of the same
  equation, which is $O(b^2)$. To put it in other words, assuming
  that $R$ is an analytic function of the coordinates, its expansion
  in powers of $b$ must be bounded term by term.
 A straightforward calculation yields
\begin{eqnarray}
R^{(1)} &=&{\cal F}_0\, f''(\rho)
        +{{\cal F}_1\,  f(\rho)+{\cal F}_2\, f'(\rho)\over (1-m/2\rho)}
\nonumber \\
  &+& {{\cal U}_0 \, u(\rho) \over (1-m/2\rho)^4}
    +{{\cal U}_1 \, u'(\rho) \over (1-m/2\rho)^3}
    +{{\cal U}_2 \, u''(\rho) \over (1-m/2\rho)^2}
\end{eqnarray}
where $'=d/d\rho$ and ${\cal F}_i$ and ${\cal U}_j$ are known
functions of $\rho$, finite at $\rho=\rho^{(B)}_{\cal H}$. Here
$g^{(B)}_{\alpha\beta}$ is taken as in Eq.~(\ref{metric}). An
examination of the expression above confirms that for $R$ to be
finite on the horizon, $f(\rho)$ must vanish on the horizon at
least as fast as $(1-m/2\rho)^2$ and $u(\rho)$ must vanish at
least as fast as $(1-m/2\rho)^4$. What does this suggest for the
correction to the horizon area formula, Eq.~(\ref{A})?

Using the metric Eq.~(\ref{metric}) we work out the expression for
$h$ with the subsequent simple result
\begin{equation}
h={(1+m/2\rho)^2\over (1-m/2\rho)^2}u(\rho)+{3f(\rho)\over
(1+m/2\rho)^{4}}.
\end{equation}
Considering the fact that we are really interested in the value of
$h$ on the horizon at its background position, namely
$\rho=\rho^{(B)}_{\cal H}$, we are faced with the observation that
$h$ {\it vanishes} on the horizon! Hence the area formula is left
unperturbed in linear approximation: any corrections to $A$ must
be of higher order, like $O(b^4)$, {\it etc.\/}  Hence by
Eq.(\ref{area}) (with $q=j=0$)
\begin{equation}
A =16\pi m^2+ O(b^4/m^2) + O(\mu^2)
\label{area2}
\end{equation}
where we have included all possible second order terms of the
correct dimensions.

The descent of the object, if sufficient slow, is known to be an
adiabatic process which causes no change in the horizon
area\cite{disturbing}.  It follows that to the stated accuracy,
$m$ is unchanged in the course of the lowering process because $A$
is preserved. When the object is finally dropped, and absorbed by
the black hole, $m$ increases by ${\cal E}$; after the suspension
machinery has been adiabatically retrieved, we acquire an
unperturbed Schwarzschild black hole with mass $m+{\cal E}$.
Calculating its horizon area from Eq.~(\ref{area2}) and
subtracting the area of what was an unperturbed Schwarzschild
black hole of mass $m$, we find the change
\begin{equation}
\Delta A=32\pi m {\cal E}+O({\cal E}^2)+\cdots\, .
\label{DeltaA}
\end{equation}
Finally substitution of Eq.(\ref{energyfinal}) gives
\begin{equation}
\Delta A\geq 8\pi\mu R  \left[1+O\left({\mu R\over
m^2}\right)\right] +\cdots.
 \label{dA}
\end{equation}
Notice that the black hole parameter $m$ has dropped out from the
dominant terms, in analogy with results for uncharged
objects\cite{BHentropy}. The minimum change in black hole entropy,
$\Delta A/4\hbar$ with the equal sign, is thus a property of the
object itself. The entropy of the object cannot exceed this
amount, lest the overall entropy of the world decrease upon the
object's assimilation (see \cite{Newbound} for the irrelevancy of
buoyancy corrections in this connection). We thus find the bound
on the entropy of an object of scalar charge $b$, characteristic
size $R$ and proper energy $E=\mu$ to coincide with Bekenstein's
proposal Eq.~(\ref{firstbound}). It wasn't improved by knowledge
of the object's scalar charge.

Actually this result can be easily generalized to the case of an
object with mass $\mu$, endowed with a {\it scalar} charge $b$ and an
{\it electromagnetic} charge $e$, which is assimilated by a RN  black hole with
charge $q$. In order to include the electromagnetic effects one should merely add
to the action functional (\ref{action}) an interaction term of the form
\begin{equation}
S_{Int}=e\int \hat A_\alpha \,\dot x^\alpha d\tau
\label{lagrangian1}
\end{equation}
where, as in Sec. III, $x^\alpha(\tau)$ denotes the particle's
trajectory, $\tau$ the proper time, an overdot stands for $d/d\tau$,
and $\hat A_\alpha$ means the electromagnetic 4--potential with the
self--field of the particle subtracted off and then evaluated at the
particle's spacetime position. The energy is now
\begin{equation}
{\cal E}= -(\mu +b\Phi) g_{0\beta}\,\dot x^\beta -e\left( {\hat
A}_0^{(q)}+{1\over 2}A_0^{(e)}\right). \label{energ_g}
\end{equation}
Here ${\hat A}_0^{(q)}$, linear in $q$, is produced by the black
hole and $A_0^{(e)}$, whose source is the object itself, is linear
in $e$. The factor ${1\over 2}$ takes care of the fact that the
object owes part of its energy to its own field, not to the
background one. As before, ${\cal E}$ corresponds to the usual
notion of energy as measured at infinity.

We require that  $q$, $e$ and $b$ be very small on the scale of $m$,
the mass of the hole. Then, correct to $O(e)$ which we regard as the
same as $O(b)$ and $O(q)$, the metric may be taken as Schwarzschild's.
Retracing the steps of the
derivation in the previous section we find that the energy of the
object at a proper distance equal to the object's proper radius $R$ is
\begin{equation}
{\cal E}\geq {2\mu R+e^2+4eq\over 8m} .
\label{energyfinal_g}
\end{equation}
The scalar field parameter is again missing due the vanishing
of the scalar self-energy.

Thus, we can declare that the entropy bound (\ref{optimalbound})
is left intact, not least due to the fact that the Birkhoff's
theorem applies in the case of the electromagnetic field. Coupling
of the scalar field to  the field generated by the electric
charges, should not open a loophole in the above claim. This is
because the corrections to the electromagnetic and scalar fields
due to the scalar-electromagnetic interaction are of second order
in the coupling constant. Therefore, if we take that to be of the
same order of magnitude as $O(b)$, then these corrections  would
induce corrections to the area formula of $O(b^4)$, which are duly
neglected. Moreover, adding a mass term to the free scalar field
action, should leave the entropy bound unaltered, provided
the Compton wavelength of the scalar field is large on the  scale of
the black hole. Accordingly, terms in the scalar field
equation, proportional to the mass of the field, can be neglected.

\section{Speculation: Sources of massive vector fields and the optimal
entropy bound}

What about sources of massive vector fields? It turns out that
most of the results that were obtained for the massless scalar and
vector fields may be used in this case. Vilenkin \cite{Vilenkin}
points out that if instead of the electromagnetic field, the
particle is coupled to a vector-meson field of vanishingly small,
but nonzero, mass, then it can be shown that the self force has
the same magnitude but opposite direction. This sharp difference
between massive and massless vector fields is a result of
different boundary conditions at the horizon surface. The basic
idea is as follows.

If the mass of the vector field $A_\nu$ is not exactly zero, then
Maxwell equations have to be replaced by Proca's:
\begin{equation}
{F^{\alpha\beta}}_{;\beta}-{\sf m}^2 A^\alpha=4\pi j^\alpha.
\label{Proca}
\end {equation}
Here and henceforth $F_{\alpha\beta}=A_{\beta ;\alpha}-A_{\alpha
;\beta}$. We assume that the mass of the vector field is very
small, namely ${\sf m}^{-1}$, the Compton wavelength of the
massive vector field, is much larger than any characteristic
distance in the problem. Therefore correct to $O({\sf m}^2 m^2)$,
the mass term in the equation above can be neglected. In solving,
let $b$ be the strength of a charge of the massive vector field
$A_\nu$ at a distance $a$ from the black hole, and let $m/2\ll{\sf
m}^{-1}$. What boundary conditions must be fulfilled for the
consistency of the solution? We require that invariants associated
with the energy-momentum tensor of the field $A_\nu$
\begin{equation}
T_\sigma^\nu={1\over 4\pi}\left( F_{\sigma\alpha}F^{\nu\alpha}
-{1\over 4}\delta_\sigma^\nu F^{\alpha\beta} F_{\alpha\beta} +{\sf
m}^2\left(A_\sigma A^\nu-{1\over 2} \delta^\nu_\sigma A_\alpha
A^\alpha\right)\right),
\label{Proca_emt}
\end{equation}
be nonsingular at the horizon, any divergence in these would
induce divergences in the invariants of the geometry via
Einstein's equations. The case in point would be that $T$, the
trace of the energy-momentum (\ref{Proca_emt}), which is
proportional to the invariant $A_\alpha A^\alpha$, must be bounded
everywhere, and the potential $A_0$ must vanish at least like
$(\rho-m/2)$ as $\rho\rightarrow m/2$. Thus all the physically
meaningful solutions of the Proca  field equation (\ref{Proca})
must satisfy the boundary condition $A_0(\rho=m/2)=0$. In the case
of a massless field, the divergence of   $A_\alpha A^\alpha$ at
the horizon causes no difficulties as long as the invariant $
F_{\alpha\beta} F^{\alpha\beta}$ is finite. This is easily seen
from Eq.~(\ref{Proca_emt}) with ${\sf m}=0$.  As usual,  we take
the charge $b$ to be a small parameter in the problem. Then the
energy-momentum tensor (\ref{Proca_emt}) is $O(b^2)$.  Hence the
same arguments we used in the massless scalar field can be used
here to show that the horizon area formula for the Schwarzschild
black hole is preserved in linear perturbation theory.

Now, since we take $m/2\ll{\sf m}^{-1}$, the field of the {\it
massive} vector field can be approximated by the solution of the
{\it massless} vector field equation. The massless vector field
equation was solved many years ago by Copson \cite{Copson} who
calculated the full electromagnetic 4-potential due to a
stationary point charge in the background of a spherical black
hole. Making use of this result with the additional requirement
that the zeroth component of the vector field vanish on the
horizon, we corroborate Linet and Leaute \cite{Leaute} by following the procedure used in
\cite{BekMayo2} and in the previous sections, to calculate the self-energy of the massive
vector field with the simple result
\begin{equation}
 {1\over 2} b A_0^{(b)}= {b^2 \over  a (1 +m/(2a))^4} {m\over 2
 a}.
\label{vector_s_E}
\end{equation}
The factor ${1\over 2}$ takes care of the fact that the object
owes part of its energy to its own field. As given earlier by
Vilenkin for $m/2\ll a$ and by Linet and Leaute  for all $a$, this self
energy has the same magnitude as in the case of the
electromagnetic field, but opposite sign. Electric charges are
{\it repelled} from neutral black holes, while the charges of
massive vector fields are {\it attracted} to them.
 The implication of this for the issue of entropy bounds is of
great importance. A straightforward calculation shows that for
constant $\rho>m/2$ and
$a\rightarrow m/2$, $A_0\rightarrow 0$, namely, as the charge is
assimilated, the massive vector field outside the black hole vanishes!
This, of course, harmonize with the No-Hair theorem. Since the
particle-vector field interaction
action is identical to the interaction action given in
Eq.~(\ref{lagrangian1}) and based on the result (\ref{vector_s_E}),
the minimal assimilation energy for
particles coupled to massive vector fields corresponds to the
equal sign in Eq.~(\ref{energyfinal_g}) with the replacement
$e^2\rightarrow -b^2$ and $q\rightarrow 0$. Correct to $O(b^2)$,
the minimal horizon area growth is given by Eq.~(\ref{DeltaA}).
Substitution of ${\cal E}$ and dividing by $4\hbar$ gives for the
entropy of the object.
\begin{equation}
 S\leq 2\pi {E R-b^2/2\over\hbar}
 \label{Sb}
\end{equation}
which is precisely the entropy bound (\ref{optimalbound}) with
$e^2\rightarrow b^2$ and $s=0$. Therefore, in a sense, the entropy
bound  (\ref{optimalbound})  was generalized to include
vector-meson charge $b$ in the same way that it was generalized to
include magnetic monopole charge $g$, $e^2\rightarrow
e^2+g^2+b^2$. This generalization, however, does not pose any
difficulty from a black-hole entropy point of
view, since black holes don't posses this quantum number.

The mass of the vector-meson field, ${\sf m}$, play an important
role in the validity of the refined entropy bound (\ref{Sb}). As
indicated before, the bound (\ref{Sb}) is correct to $O({\sf m}^2
m^2)$. So if for example we consider the $\rho$ vector-meson
(${\sf m}= 770\, MeV$), then the mass of the black hole must be
smaller than $m_{Pl}^2/{\sf m}\simeq 10^{15}\, {\em gr}$, the mass
range for mini black holes. However, if ${\sf m}$ is large then
the mass term in Proca's equations (\ref{Proca}) cannot be
neglected. Nevertheless, the field generated by $b$ is now a short
range field. Although there is now a contribution to the
energy-momentum tensor from this field, it is localized around the
object, and thus can be lumped into its usual energy-momentum
tensor. No novel perturbation to the metric arises from this.
Hence, $b$ cannot directly perturb the horizon area formula
(\ref{area}), and so $m$ is unaffected by slow lowering of the
object.  Furthermore, no novel potential term is contributed to
${\cal E}$ by the field unless the particle is already next to the
horizon; otherwise the short range field does not reach down to
the horizon and cannot polarize it.  Hence the change in horizon
area turns out to be $b$--independent, and $b$ cannot appear in a
generic entropy bound.

We conclude that the conjecture, that the entropy bound
\begin{equation}
S\leq 2\pi {\sqrt{E^2 R^2-s^2}-e^2/2\over\hbar}
\end{equation}
for an object with spin $s$, charge $e$, maximal radius $R$ and
mass-energy $E=\mu$ is the tightest {\it generic} bound on
entropy, seems reasonable.

{\bf ACKNOWLEDGMENTS} The author thanks Professor J. D. Bekenstein
for his suggestions and advice, to Shahar Hod for discussions and
to Bernard Linet and Lior Burko for correspondence. This research
is supported by a grant from the Israel Science Foundation,
established by the Israel Academy of Sciences and Humanities.

\appendix
\section{ Field of scalar charge in black hole background}

Here we determine $\Phi$ resulting from a scalar charge $b$ in the
Schwarzschild background Eq.~(\ref{metric}). Using the conventions
of Misner, Thorne and Wheeler\cite{MTW} we write the
Klien--Gordon equation for the axisymmetric stationary massless
scalar field of a test point scalar charge $b$ situated at
$\{\rho,\theta\}=\{a,0\}$ as
\begin{equation}
(1-(m/2\rho)^2)\Delta\Phi+(m^2/2\rho^2)
\overrightarrow\rho\cdot\overrightarrow\nabla\Phi=-4\pi b\delta(\rho-a)\delta(\theta)\delta(\phi).
\label{Laplace}
\end{equation}
Here $\Delta$ and $\overrightarrow\nabla$ are the usual Laplacian
and Gradient operators in flat spacetime respectively.
 The potential $\Phi$ of the scalar charge $b$ may be looked for in the
 form\cite{Hadamard}
\begin{equation}
\Gamma_a^{-1/2}\left[U_0+U_1\,\Gamma_a+U_2\,\Gamma_a^2+\ldots\right].
\label{anzats}
\end{equation}
$\Gamma_a$ here denotes the square of the geodesic distance from
the source location in the space whose metric is (\ref{metric}),
namely $\Gamma_a\equiv\rho^2+a^2-2\rho\, a\cos\,\theta$. $U_0,\,
U_1, \, U_2,\ldots$ are analytic functions of $\rho$ for
$\rho>m/2$. Let us scale $\rho$ by the rule
 $\rho\rightarrow 2\rho/m$. Substituting the elementary
 solution (\ref{anzats}) in Eq.~(\ref{Laplace}) and analyzing
 the first three terms suggests that instead of determining
 successively the remaining $U_n$, we should look for the form
\begin{equation}
\Phi=B {\rho\over \rho^2-1}F(\gamma), \quad\quad \gamma\equiv
\Gamma_a/(\rho^2-1)\label{F}
\end{equation}
where $B$ is a constant to be determind later. Doing so, we observe that $F(\gamma)$ obeys
\begin{equation}
2\gamma(\gamma+a^2 -1){d^2\over d \gamma^2}
F(\gamma)+3(2\gamma+a^2-1){d\over d \gamma} F(\gamma)+2
F(\gamma)=0.
\label{Feq}
\end{equation}
Therefore the solution of Eq.~(\ref{Feq}) is a linear combination
of \cite{Mathematica}
\begin{eqnarray}
&& F_1={1\over \sqrt\gamma\sqrt{\gamma+a^2 -1}},
\nonumber \\
&& F_2={1\over \sqrt\gamma\sqrt{\gamma+a^2 -1}}
\log\left({\sqrt\gamma +\sqrt{\gamma+a^2 -1}\over \sqrt{a^2-1}}\right).
\label{F1F2}
\end{eqnarray}
Substituting for $\gamma$ and $F$ in the definition (\ref{F}) and
rescaling $\rho\leftarrow 2\rho/m$, we find for $\Phi$
\begin{eqnarray}
&& \Phi_1=B_1{\rho\over
\sqrt{\Gamma_a\,\Gamma_{\tilde{a}}}}, \nonumber \\
 && \Phi_2= B_2{\rho\over
\sqrt{\Gamma_a\,\Gamma_{\tilde{a}}}}\, \log\,\left({
{\sqrt{\Gamma_a}}+\sqrt{\Gamma_{\tilde{a}}} \over \sqrt{1-(m/ 2
\rho)^2} (2\rho /m)(2 a/ m)\sqrt{1-(m/ 2 a)^2}} \right)
\end{eqnarray}
where $\tilde{a}\equiv m^2/4 a$.

Our approach is perturbative in nature. Physical invariants that
may be assembled from the energy-momentum of the perturbation must
thus be bounded everywhere, including at the horizon: any
divergence would imply divergence of the curvature invariants. As
can be easily verified, every invariant of the geometry associated
with this solution is proportional to   $\left(\Phi_{,\alpha}
\Phi^{,\alpha}\right)^k$. Now, since $\Phi_{,0}$ is assumed to be
identically zero and since we are using the metric (\ref{metric}),
the mentioned invariant would be bounded provided the solution and
its gradient are bounded everywhere. Now, both solutions $\Phi_1$
and $\Phi_2$ vanish at spatial infinity; for $m/2< a<\infty$ and
$\rho\rightarrow\infty$ they are $O(1/\rho)$. Furthermore, both
solutions are singular at the charge location,
$\{\rho,\theta\}=\{a,0\}$ as required. However, for $a>m/2$,
$\Phi_2$ diverges logarithmally everywhere on the horizon
$\{\rho=m/2,\forall\theta\}$. We thus reject it as a physical
solution.

The remaining solution $\Phi_1$ has some intriguing
characteristics. Firstly, for constant $\rho>m/2$ and
$a\rightarrow m/2$, $\Phi\rightarrow 0$, namely, as the charge is
assimilated, the scalar field outside the black hole vanishes! (see
Eq.~(\ref{Final_Phi}) below). This, of course, agrees with the
No-Hair theorem for black
holes. Secondly, for constant $a>m/2$ and $\rho\rightarrow m/2$
the value of the scalar field is finite. It is true that in the
limit $a\rightarrow m/2$, $\rho\rightarrow m/2$ and
$\theta\rightarrow 0$ $\Phi_1$ diverges. But, one should not be
alarmed by this, since this divergence is localized at the point
where the scalar charge touches the horizon,
$\{\rho,\theta\}=\{a=m/2,0\}$ and does not encompass the whole of
the horizon. Furthermore, this divergence can be attributed to our
neglect of  the self-energy of the particle.

We thus infer that the elementary solution of Eq.~(\ref{Laplace})
with source at $\{\rho,\theta\}=\{a,0\}$ is
\begin{equation}
\Phi(\rho,\theta)=-b\,{1-(m/2a)\over 1+(m/2a)}{\rho\over
\sqrt{\Gamma_a\,\Gamma_{\tilde{a}}}}
\label{Final_Phi}
\end{equation}
in which the constant $B_1$ was  set by the asymptotic value of the
field at the position of the charge.

The expression for the scalar field $\Phi(\rho,\theta,\phi)$ due
to a charge $b$ situated at the point $\{\rho',\theta',\phi'\}$ can
be obtained from Eq.~(\ref{Final_Phi}) by a rotation of the axes,
 which manifests itself by the simple replacement
\begin{equation}
\cos\,\theta\rightarrow\chi(\theta,\phi ; \theta'\phi')=\cos\,\theta\,\cos\,\theta'+
\sin\,\theta\,\sin\,\theta'\cos(\phi-\phi').
\end{equation}
The result is analogous to the one  in Eq.~(\ref{Final_Phi}) with
$\Gamma_{\rho'}\rightarrow\Gamma_{\rho',\theta',\phi'}\equiv\rho^2+\rho'^2-2\rho
  \, \rho'\,\chi(\theta,\phi ;\theta'\phi')$. This provide the means to calculate the scalar
 field originating from any arrangement of scalar charges by means of the
following formula
\begin{equation}
\Psi\,(\rho,\theta,\phi)=\int\,\Phi(\rho,\theta,\phi ;
\rho',\theta',\phi')
\Sigma(\rho',\theta',\phi')\,\sqrt{-g}\,d^3 x'
\label{general}
\end{equation}
where $\Sigma(\rho',\theta',\phi')$ is the charge distribution
density of a specified scalar charges configuration and the
integration is assumed over a constant time slice of the
spacetime.

%\begin{references}

{}


\begin{thebibliography}{}
\bibitem{Mathematica} This solution can be easily found by the use
of any cumputer enviroment for doing mathematics such as
Mathematica.
\end{thebibliography}
\end{document}